\documentclass[njp,12pt]{iopart}
\usepackage{graphicx}
\expandafter\let\csname equation*\endcsname\relax
\expandafter\let\csname endequation*\endcsname\relax
\usepackage{amsmath}
\usepackage{amsfonts}
\usepackage{amssymb}
\usepackage{soul}
\usepackage{dsfont}
\usepackage{hyperref}
\usepackage{amstext}
\usepackage{braket}
\usepackage{color,soul}
\usepackage{dcolumn}
\usepackage{bm}
\usepackage{tabularx,ragged2e,booktabs,caption}
\usepackage{cellspace}

\usepackage{floatrow}
\usepackage{hyperref}
\hypersetup{colorlinks=true, citecolor=blue, urlcolor=blue, linkcolor=blue}
\usepackage{color,soul}
\usepackage[a4paper, total={6.5in, 9in}]{geometry}

\date{\today}
\newcommand{\be}{\begin{equation}}
\newcommand{\ee}{\end{equation}}
\newcommand{\bea}{\begin{eqnarray}}
\newcommand{\eea}{\end{eqnarray}}

\begin{document}
\title{Entanglement of quantum oscillators coupled to different heat baths}


\author{Wei-Can Syu, Da-Shin Lee,
Chen-Pin Yeh}
\address{ Department of Physics, National Dong Hwa University, Hualien 97401, Taiwan, R.O.C.}
\ead{syuweican@gmail.com}

\begin{abstract}
We study the non-equilibrium dynamics of two coupled oscillators
interacting with their own  heat
baths of quantum scalar fields at different temperature $T_1$ and $T_2$ with bilinear
couplings between them. We particularly focus
on the entanglement or inseparability property of their quantum
states.   The critical temperatures of two respective oscillators, $T_{1c}$ and $T_{2c}$, higher
than which the entanglement disappears, can  be determined. It
is found that when two damping
parameters are largely different, say $\gamma_1 \ll \gamma_2$,
the critical temperature $T_{1c}$ with respect to the frequency $\Omega_+$, the higher frequency among two normal modes frequencies, can be
very large,  $T_{1c} \gg \Omega_+$, while $T_{2c} \propto \Omega_+$ with the possibility of hot entanglement. The entanglement of two oscillators with the temperature-dependent damping parameters $\gamma_{1;2,T}$ from heat baths is also discussed.
\end{abstract}
\maketitle
\newpage
\section{Introduction}
Entanglement or inseparability of composite quantum systems is a
characteristic trait of quantum mechanics lacking in classical
mechanics, and then becomes a key resource in many quantum
information processing protocols. The degree of entanglement
or mixture is the central concern from the
viewpoints of quantum information and quantum computation. Thus,
it is of great importance to search for a proper mathematical
frameworks to quantify such features in general mixed quantum
states. A step toward this goal is to find the criterions to
distinguish between quantum and classical correlations of mixed
states. A {bipartite} state is said to be separable if the
total density matrix $\rho_{AB}$ can be written in the form,
$\rho_{AB}=\sum_i p_i\left(|\phi_i\rangle\langle\phi_i|\right)_A
\otimes \left(|\psi_i\rangle\langle\psi_i|\right)_B$  with
$p_i>0$. Two subsystems $A$ and $B$ are said to be quantum
entangled if they are not separable. One criterion of the
separability  was proposed by Peres~\cite{Per} and later further
extended by Horodecki~\cite{Hor,Hor2} in the case of discrete variable
quantum systems in finite-dimensional Hilbert space. The
application of Peres-Horodecki criterions to continuous variable
systems in infinite-dimensional Hilbert space~\cite{Simon} has
also received much attention. Some other entanglement measures
such as negativity~\cite{Vid} and logarithmic
negativity~\cite{Ple,Eis,Vir} are also suggested for general
{bipartite}
states.

The quantum-mechanical degrees of freedom unavoidably couple to environments or baths, which induce dissipation and
decoherence.  The effective theory can be described by the reduced
density matrix, which is obtained by tracing out  environmental
variables in the full density matrix using the method of
Feyman-Vernon influence functional \cite{Fey}, and becomes an
essential quantity for constructing the entanglement measures
mentioned above.  This concept has been used in pioneering works
on quantum Brownian motion and general open quantum systems
\cite{Cal,Cal2,Cal3}. {Recently} there have been remarkable experiments
\cite{Gro1,Tho,Gro2,Nai} using an opto-mechanical resonator to
engineer the properties of the environmental degrees of freedom
and its coupling to the probed systems, which make it possible to
couple with several different baths, and to tailor decoherence and
dissipative properties in such an experimentally controlling
manner. {Thus, the experimental advances motivate theoretical
studies, showing that coupling to different non-equilibrium
environments can lead to the persistence of entanglement in the
high temperature limit~\cite{Gal,Est,Ved}. The idea is to consider
the model of two coupled, parametrically driven, dissipative
harmonic oscillators with the typical energy scale $E$, which
couple  to its own heat bath at the same temperature $T$. {It is
expected that} when $k_B T
> E$, the quantum nature of the system such as superposition and
entanglement will be lost. {However,} it was
argued that the nonequilibrium dynamics can bring the system into
the late-time saturation
characterized by an effective temperature, which is the relevant
scale for determining whether or not quantum features of the
system can survive~\cite{Ved}. The effective temperature can be
much lower than the temperature of the bath $T$ so
as to sustain quantum entanglement even at high temperature
$T$~\cite{Ved}. The works of~\cite{Hsiang2,Hsiang3} reexamine the
above-mentioned model by considering  time-independent
couplings between two oscillators with the same Ohmic
damping parameters, $\gamma_1=\gamma_2$. Additionally  the
temperatures of two baths are set to be different, say $T_1$ and
$T_2$ respectively. The late-time nonequilibrium steady states are
found where the critical temperature below which quantum
entanglement exists turns out not so surprisingly to be $k_B T_{c}
\sim E \sim \Omega_+$ where  $\Omega_+$ is the  higher frequency among two normal modes frequencies. It seems that parametric driving with the
time-dependent coupling between two oscillators is a key mechanism
for sustaining the entanglement in the system at high temperature.
Nevertheless,  in \cite{Boy}, coherence of two mechanical
oscillators coupled to different and uncorrelated baths is
studied, and is found to be enhanced in the case of $\gamma_1 \neq
\gamma_2$. Thus, in this work we plan to extend the works of
\cite{Hsiang2,Hsiang3} by considering the case when $\gamma_1 \neq
\gamma_2$ with the hope to boost the critical temperature for
the entanglement to a higher one as compared with the case of
$\gamma_1=\gamma_2$. Recently, the model above has also been adopted
to successfully explain the experimental demonstration of a novel
effect of heat transfer between two objects driven by quantum
vacuum fluctuations~\cite{Fon}. Theoretic studies of this model on heat transfer between microsystems are also found for example in ~\cite{Meher,Biehs,Xuereb,Asadian}.

Our presentation is organized as follows. In next section, we
introduce the model, and briefly review the idea of open quantum
system with the method of the closed-time-path formalism. The
environmental degrees of freedom in the full density matrix of the
system-plus-environment are traced over to obtain the reduced
density matrix of the system. Environmental effects are then all
encoded in the influence functional, which can be determined by
nonequilibrium two-point correlators. Then the corresponding
Heisenberg-Langevin equations for two oscillators involving the damping as well as noise terms are derived, which can
be solved for the general solutions. In Sec. \ref{sec3}, we
construct the covariance matrix, and substitute the solutions of
the Heisenberg-Langevin equations into the matrix elements. In the
late-time limit, the matrix elements are dominated by
the noise correlations on the oscillators. The separability
criterions are discussed. Violation of the criterions allows us to
compute the critical temperature of the entanglement both
numerically and analytically in Sec. \ref{sec4}. We summarize all
results with their implications in Sec. \ref{sec5}.

\section{The model: two oscillators coupled to its own heat baths}\label{sec2}

We consider the model of two quantum oscillators coupled to their
own heat baths of quantum scalar fields at different temperatures.
The action is described
by~\cite{Hsiang1,Hsiang2,Hsiang3} {\bea \label{action}
S[ \chi, \varphi] &=& S_{\chi}[\chi_1,\chi_2]+ S_{\varphi}[\varphi_1,\varphi_2]+ S_{\chi\varphi}[\chi_{1,2},\varphi_{1,2}] \nonumber\\
&=& \int \, ds \, \bigg[ \frac{m}{2} (\dot\chi_1)^2 -\frac{m \Omega^2}{2} \chi_1^2 +  \frac{m}{2} (\dot\chi_2)^2 -\frac{m \Omega^2}{2} \chi_2^2  -m\sigma \chi_1(s) \chi_2(s) \bigg] \nonumber\\
&&\quad\quad  + \int d^4 x\, \frac{1}{2} \bigg[ \partial_\mu \varphi_1 \partial^\mu \varphi_1 + \partial_\mu \varphi_2 \partial^\mu \varphi_2 \bigg]\nonumber\\
&&\quad\quad \quad+  g_1 \int d^4 x \, \chi_1 (s)\, \delta^{(3)} ({\bf{x}}- {\bf{z}}_1 (s)) \, \varphi_1(x) \nonumber\\
&&\quad\quad\quad +  g_2 \int d^4 x \, \chi_2 (s)\, \delta^{(3)}
({\bf{x}}- {\bf{z}}_2 (s)) \, \varphi_2(x) \, . \eea} Two
oscillators have the same oscillation frequency $\Omega$ and  mass $m$, and
they couple to each other with a coupling constant $\sigma$.  The
coupling of the oscillator $\chi_{1,2}$ to the heat bath of a
scalar field $\varphi_{1,2}$ has the coupling strength $g_{1,2}$.
Moreover, ${\bf z}_{1,2}$ is the prescribed trajectory of
the oscillators. Here we assume that the initial density matrix
for the whole system (oscillators-plus-fields) at
time $t_i =0$ is factorizable as
   \be \rho(0)=\rho_{\chi}(0)
\otimes \rho_{\varphi_1} \otimes \rho_{\varphi_2} \, .
   \ee
The fields are initially in thermal equilibrium at temperature
$\beta_1=1/T_1$ for the field $\varphi_1$ and $\beta_2=1/T_2$ for
the field $\varphi_2$ with $T_1\neq T_2$ in general. Their
respective density matrices are
 given by
\begin{equation}\label{initialcondphi}
    \rho_{\varphi_{1}}=e^{-\beta_1 H[\varphi_1]}/ {\rm Tr} \{ e^{-\beta_1 H[\varphi_1]}  \} \,, \quad\quad \rho_{\varphi_{2}}=e^{-\beta_2 H[\varphi_2]}/ {\rm Tr} \{  e^{-\beta_2 H[\varphi_2]}\} \,,
\end{equation}
where $H [\varphi_{1,2}]$ is the Hamiltonian for the free field
$\varphi_{1,2}$, constructed from $S_\varphi
[\varphi_1,\varphi_2] $ in (\ref{action}). The  density matrix
$\rho (t)$ of the whole system  evolves unitarily according to \be
\rho (t) = U(t,0 ) \, \rho (0) \, U^{-1} (t ,0)\,, \ee where $U(t
,0 )$ is the time evolution operator. The effects from the
environment to the system can be summarized in the reduced density
matrix $\rho_r (t)$, which is obtained by tracing out the
environmental degrees of freedom in $\rho (t)$. Here we provide a
brief summary of the main results of the reduced density matrix of
the system, and a mini review of how to obtain the equations of
motion for the position operators of two coupled oscillators with
environmental effects. We start with the reduced density matrix,
expressed as
\begin{align}
&\quad\rho_{\chi}(\chi_{1 F},\chi'_{1 F}, \chi_{2 F},\chi'_{2 F};t) =\int_{-\infty}^{\infty}\!\left\{\prod_{a=1}^{2}d\chi_{a I} d\chi'_{aI}\right\}\;\rho_{\chi}(\chi_{1I},\chi'_{1I},\chi_{2I},\chi'_{2I},0)\nonumber\\
&\quad \left\{\prod_{a=1}^{2}\int_{\chi_{aI}}^{\chi_{aF}}\!\mathcal{D}\chi_{a+}\!\int_{\chi'_{aI}}^{\chi'_{aF}}\!\mathcal{D}\chi_{a-}\right\}\exp\Bigl(i\,S_{\chi}[\chi_{1+}, \chi_{2+}]-i\,S_{\chi}[\chi_{1-},\chi_{2-}]\Bigr)\times \mathcal{F}[\chi_{1+},\chi_{2+},\chi_{1-},\chi_{2-}] \, .\notag
\end{align}
The influence functional $\mathcal{F}$  can be written in terms of nonequilibrium two-point
correlators constructed from the environment fields as
 \begin{align}
&\quad\mathcal{F}[\chi_{1+},\chi_{2+},\chi_{1-},\chi_{2-}]
=e^{i\,S_{IF}[\chi_{1+},\chi_{2+},\chi_{1-},\chi_{2-}]}\notag\\
&\quad\quad
=\prod_{a=1}^{2}\exp\biggl\{\frac{i}{2}\,\int_{0}^{t}\!ds\,ds'\biggl(\Bigl[\chi_{a+}(s)-\chi_{a-}(s)\Bigr]G_{R,\,T_a}(s,s')\Bigl[\chi_{a+}(s')+\chi_{a-}(s')\Bigr]\biggr.\biggr.\notag\\
&\qquad\qquad+\biggl.\biggl.i\,\Bigl[\chi_{a+}(s)-\chi_{a-}(s)\Bigr]G_{H,\,T_{a}}(s,s')\Bigl[\chi_{a+}(s')-\chi_{a-}(s')\Bigr]\biggr)\biggr\}\,,
 \end{align}
where $S_{IF}$ is called the influence action. The retarded
Green's function $G_{R,\,T_a}$ is defined by \be
G_{R,\,T_a}(s,s')=i \, g_{a}^2 \,\theta(s-s')\, \Big\langle
\Bigl[\varphi_{a}({\bf z}_{a}(s),s),\varphi_{a}({\bf
z}_{a}(s'),s')\Bigr]\Big\rangle \ee {and the {Hadamard function}
$G_{H,\,T_{a}}$ by} \be  G_{H,\,T_{a}}(s,s')=\frac{g_{a}^2}{2}\,
\Big\langle \Bigl\{\varphi_{a}({\bf z}_{a}(s),s),\varphi_{a}({\bf
z}_{a}(s'),s')\Bigr\}\Big\rangle\, \ee with $a=1,2$. Notice that
we have absorbed the coupling constant $g_{a}^2$ into the
definition of the Green's functions. The Hadamard function is
simply the expectation value of the anti-commutator of the quantum
field $\varphi_a$ with respect to the thermal state at temperature
$T_a$, and notice that the retarded Green's function does not have
any temperature dependence due to the linear coupling of the
scalar field to the oscillator. The kernels $G_{H,\,T_{a}}$ and
$G_{R,\,T_a}$ respectively are in turn linked by the
fluctuation-dissipation relation. The fluctuation-dissipation
relation is known to play a pivotal role in balancing these two
effects in order to dynamically stabilize the nonequilibrium
evolution of the system under a fluctuating environment.
Mathematically, it relates the Fourier transform of the
fluctuation kernel  $G_{H,\,T_{a}}$  to the imaginary part of the
retarded kernel $G_{R,\,T_a}$ as follows \bea
&&G ( s,s')= \int \frac{d \omega }{2\pi} \,G ({\bf z}(s),{\bf z}(s'); \omega) \, e^{-i \omega (s-s')}\,,\nonumber\\
&&G_{H,T_a}({\bf z}(s),{\bf z}(s'); \omega) =
\coth\bigg[\frac{\omega}{2 T_a} \bigg] \,  {\rm Im}
G_{R,\,T_a}({\bf z}(s),{\bf z}(s'); \omega)  \, . \eea

Thus, one can define the coarse-grained effective action as
\be
S_{CG}[\chi_{1+},\chi_{2+},\chi_{1-},\chi_{2-}]  =S_{\chi}[\chi_{1+}, \chi_{2+}]-S_{\chi}[\chi_{1-}, \chi_{2-}]+ S_{IF}[\chi_{1+},\chi_{2+},\chi_{1-},\chi_{2-}] \,.
\ee
To find the time evolution of  two oscillators, we derive the Heisenberg-Langevin
equations for the quantum operators $\hat\chi_1$ and $\hat\chi_2$ that
incorporate  environmental effects. This can be done
by introducing an auxiliary variable $ \eta_{T_a} (s)$, the noise
force, with a Gaussian distribution function:
\begin{equation}
P[\eta_{T_a} (s)] = \exp \left\{ - \frac{1}{2}  \, \int_{0}^{t} ds
\, \int_{0}^{t} ds' \, \eta_{T_a} (s) \, G^{ -1}_{H ,T_{a}} (s-s') \, \eta_{T_a}
(s') \right\} \, . \label{noisedistri}
\end{equation}
In terms of the noise force $\eta_{T_a} (s) $,  $ S_{CG} $ can be
rewritten as an ensemble average over $ \eta_{T_a} (s)$,
\begin{equation}
\exp i S_{CG}[\chi_{1+},\chi_{2+},\chi_{1-},\chi_{2-}]  =  \int \!
\prod_{a=1}^{2} \! \mathcal{D}\eta_{T_a} \, P [\eta_{T_a} (s)] \,
\exp i S_{\eta} \left[\chi_{1+},\chi_{2+},\chi_{1-},\chi_{2-} ;
\eta_1,\eta_2 \right] \, ,
\end{equation}
where the stochastic coarse-grained effective action   $ S_{\eta
}$
is given by
\bea
&& S_{\eta}\left[\chi_{1+},\chi_{2+},\chi_{1-},\chi_{2-} ; \eta_1,\eta_2 \right]  =S_{\chi}[\chi_{1+}, \chi_{2+}]-S_{\chi}[\chi_{1-}, \chi_{2-}]  \nonumber \\
&& \quad\quad\quad   + \sum_{a=1}^{2} \frac{1}{2}\,\int_{0}^{t}\!ds\,\int_0^t \, ds'\Bigl[\chi_{a+}(s)-\chi_{a-}(s)\Bigr]G_{R,\,T_a}(s,s')\Bigl[\chi_{a+}(s')+\chi_{a-}(s')\Bigr]\biggr.\biggr.\nonumber\\
&& \quad\quad \quad\quad \quad\quad \quad\quad \quad\quad
\quad\quad+  \sum_{a=1}^{2} \int_{0}^{t} ds \, \eta_{T_a} (s)
\Bigl[\chi_{a+}(s)-\chi_{a-}(s)\Bigr]\, . \eea Then, the equations
of motion for $\hat\chi_1$ and $\hat\chi_2$ can be
derived from the action $S_{\eta}$ by introducing the center of mass
coordinate $\chi_{a}$ and the relative coordinate $R_a$, $
    \chi_{a}=\frac{1}{2}\bigl(\chi_{a+}+\chi_{a -}\bigr)\,, R_a=\chi_{a +}-\chi_{a -}\, $,
and then taking the variation of $S_{\eta}$ with respect to $R_a$. For
simplicity, we assume that two oscillators barely move away from the
spatial locations of ${\bf z}_{0 a}$ so that ${\bf z}_a (s) \simeq
{\bf z}_{0 a}$. In this approximation, the retarded Green's
function $G_{R,\,T_a}$ and the Hadamard function $G_{H,\,T_a}$ can
be expressed as \bea
G_{R,\,T_a}(s,s')&=& -\frac{1}{2\pi} \,g_{a}^2\, \theta(s-s') \, \delta' (s-s')\, , \nonumber\\
G_{H,T_a} (s,s') &=& \frac{g_{a}^2}{4 \pi} \int \frac{ d \omega}{2\pi} \omega \coth \bigg[ \frac{\omega}{2 T_a} \bigg] \, e^{-i \omega (s-s')}\, .
\eea
With the Green's functions above, the Heisenberg-Langevin  equations  for quantum oscillators can be written as
\begin{align}
\ddot{\hat\chi}_1 \, + 2\gamma_{1}\, \dot{\hat\chi}_1+\Omega^2\hat\chi_1+\sigma\hat\chi_2=\frac{\eta_{T_1}}{m}, \label{chi1} \\
\ddot{\hat\chi}_2\, +2\gamma_{2}\, \dot{\hat\chi}_2+\Omega^2\hat\chi_2+\sigma\hat\chi_1=\frac{\eta_{T_2}}{m} \, . \label{chi2}
\end{align}
The effects from the retarded Green's function $G_{R a}$ not only slow down the motion of $\chi$
through the induced damping terms with the damping parameters $\gamma_a=g_a^2/8 \pi m$, but also give a shift to the
oscillation frequency, with $\delta \Omega_a^2=-4 \gamma_a \delta(0)$. Here we consider the weak oscillator-bath couplings with small $g_a$ where the corrections $\delta \Omega_a^2 $ can be ignored as compared with $\Omega^2$.

It worths mentioning that there are two different sources of
fluctuations, over which the averages are taken on  $\hat \chi_1$ and
$\hat\chi_2$. One is the average over intrinsic quantum fluctuations
of the oscillators, and the other is the average over the noise
manifested from thermal fluctuations of the environments. The
distribution function $P[\eta_{T_a}(s)]$ in (\ref{noisedistri})
leads to the correlation function of the noise as follows
\be \langle \eta_{T_a} (s) \,  \eta_{T_a} (s') \rangle= G_{H, T_a}
(s-s') \,, \ee where
\begin{align} \label{G_H}
G_{H, T_a}(s-s')= 2 \, m \gamma_{a}\int
\frac{d\omega}{2\pi}\,\omega \coth\bigg[{\frac{\omega}{2
T_a}}\bigg] e^{-i\omega(s-s')} \,.
\end{align}

In~\cite{Boy}, the model of~(\ref{action}) has been considered
where  the environment consists of a collection of harmonic
oscillators. The Heisenberg equations of motion of the system and
the environment can be solved exactly, giving the same form of the
equations in (\ref{chi1}) and (\ref{chi2}) with the Green's
functions constructed out of the harmonic oscillators.  In this
work, we derive the Heisenberg-Langevin equations  from the
influence functional.

 \section{Covariance matrix and separability criterions}\label{sec3}

The initial states of two quantum oscillators are Gaussian wave
packets with the initial conditions on the expectation values of the position operator $\hat\chi_a$
and the momentum operator $\hat p_a= m \dot{\hat\chi}_a$
 as follows}
 \bea
\langle \hat\chi_a (0) \rangle &=& \langle \hat p_a (0) \rangle=0 \, ,\quad\quad \langle \{ \hat\chi_a (0), \hat p_b (0)\}\rangle=0 \, , \label{ini_cond_1}\\
\langle \{\hat \chi_a (0), \hat \chi_b(0) \}\rangle &=&
\delta_{ab} \, \langle {\hat\chi}^2_a (0) \rangle \, , \quad\quad
\langle \{ \hat p_a (0), \hat p_b(0)\} \rangle= \delta_{ab}\,
\langle {\hat p}^2_a (0) \rangle \, , \label{ini_cond_2}
 \eea
where two oscillators are initially in a separable state. The interaction between two oscillators starts to build up the
entanglement between them.

A Gaussian state is completely characterized by its first and
second statistical moments of a {raw matrix}
defined as ${\bf \hat
X}=(\hat\chi_1,\hat p_1 ,\hat\chi_2,\hat p_2)$,
where the first moments vanish, $\langle {\bf \hat X} \rangle =0 $
due to the chosen initial conditions and the second moments are
specified by the covariance matrix given by
   \be {
V}_{ij}=\frac{1}{2} \langle  { \hat X}_i { \hat X}_j + { \hat X}_j
{\hat X}_i \rangle- \langle {\hat X}_i \rangle \langle { \hat
X}_j\rangle \,
  \ee
with $i,j=1,2,3,4$. For any operator $\hat O$,  $\langle \hat O
\rangle={{ \rm Tr} [\rho_\chi \hat O ]}$ where $\rho_{\chi}$ is
{the reduced}
density matric of $\chi_1$ and $\chi_2$. Apparently, ${\bf V}$ is
a real symmetric matrix, namely ${\bf V}={\bf V}^T$. Also, with
the Gaussian states described above, we have
$V_{23}=-V_{14}$,\,\,$V_{31}=V_{13}$,\,\,$V_{32}=-V_{41}$,
\,\,$V_{41}=-V_{14}$,\,\,$V_{42}=V_{24}$ and
$V_{12}=V_{21}=V_{34}=V_{43}=0$ where  the covariance matrix can
be written as
\begin{align}
\mathbf{V}=\begin{pmatrix}
A  & C\\
C^T & B
\end{pmatrix}
\end{align}
with
\begin{align}
A=\begin{pmatrix}
V_{11}& 0\\ 0& V_{22}
\end{pmatrix},\quad
B=\begin{pmatrix}
V_{33}& 0\\ 0& V_{44}
\end{pmatrix},\quad\text{and}\,\,\,
C=\begin{pmatrix}
V_{13}& V_{14}\\ -V_{14}& V_{24}
\end{pmatrix}.
\label{ABC}
\end{align}
Considering the {positive definite} density matrix
$\rho_{\chi}$, Heisenberg uncertainty principles  can be cast in the form
\cite{Simon1,Simon2}
  \be
  {\bf V} +i {\bf K} \ge 0 \,  \label{Hei}
  \ee
with the matrix $\bf K$ defined as
 \be
 {\bf K}=\begin{pmatrix}
0& 1 &0 & 0\\
-1& 0 &0&0\\
0&0&0&1\\
0&0&-1&0
\end{pmatrix}\,,
 \ee
and should be obeyed by any quantum system. The uncertainty principle  is a
direct consequence of the canonical commutation relation  and the
non-negativity of the density matrix. Moreover {\bf V} can be
diagonalized by the sympletic transformation $S \in Sp(4,R)$, the
ten-parameter real symplectic group, where {two} sympletic
eigenvalues, $\eta_{\lessgtr}$ {(with $\eta_> > \eta_<$)} would
satisfy $\eta_\lessgtr \ge 1/2$ as the result of (\ref{Hei}) and
encode essential information for the Gaussian states~\cite{Simon}. According to \cite{Per,Hor},
one can define the partial transpose $\bar \rho_{\chi}$ of a
bi-particle quantum state $\rho_{\chi}$ as the transpose performed
on  one of two particles in a given basis. Accordingly, the the
partial transpose covariance matrix is denoted as  ${\bf \bar V}$.
Together with the positivity of the partially transposed density
matrix $\bar\rho_{\chi}$, the necessary and sufficient
separability criterion (Peres-Horodecki-Simon (PHS)  criterion)
for {bipartite} continuous Gaussian variable systems
is then expressed as \cite{Simon}
   \be {\bf \bar V} +i {\bf K} \ge 0 \, .
   \label{PHS}
   \ee
The violation of the PHS criterion
indicates the existence of quantum entanglement between two Gaussian systems.
Again, the associated {two} sympletic eigenvalues
$\bar \eta_{\lessgtr}$ {of ${\bf \bar V}$}
would satisfy $\bar
\eta_{\lessgtr}\ge 1/2$ due to (\ref{PHS}). The
expressions of $\bar \eta_{\lessgtr}$  in terms of the matrix $A$, $B$, and $C$ defined above
will be discussed below as a measure of quantum entanglement of
the Gaussian states in this study.

To find the covariance matrix we need to solve Eqs.(\ref{chi1})
and (\ref{chi2}) where the general solutions are the sum of the
homogeneous solution $\chi_{a;h}$  that depends on the initial
conditions, $\chi_a (0)$ and $\dot\chi_a (0)$, and the particular
solution $\chi_{a;p}$ that satisfies the full equations with the
noise. Let us write the time dependence of $\chi_{a;h}  $ as
$\chi_{a;h} (t) \propto e^{-i \bar{\Omega}_{a,\pm} t}$.  The frequency
$\bar{\Omega}_{a,\pm}$ is determined by the equation, $\tilde{\bf
D}^{-1} (\bar{\Omega}_{a,\,\pm })=0$ with the matrix $\tilde{\bf D}
(\omega)$ defined as
\begin{align}\label{tildeD}
\tilde{\bf D}(\omega)&=\frac{1}{-\omega^2\mathbf{I}+\mathbf{\Omega}^2-i2\omega\mathbf{\Gamma}}\,,
\end{align}
where
\begin{align}
&\mathbf{I}=\begin{pmatrix}
1 & 0 \\
0 & 1
\end{pmatrix},\quad
\mathbf{\Omega}^2=\begin{pmatrix}
\Omega^2 & \sigma \\
\sigma & \Omega^2
\end{pmatrix},\quad
\mathbf{\Gamma}=\begin{pmatrix}
\gamma_1 & 0 \\
0 & \gamma_2
\end{pmatrix}\, .
\end{align}
This gives the following determinant,
\be
\det\tilde{D}^{-1}(\omega)=\left[(-\omega^2+\Omega^2-i2\omega\gamma_{1})(-\omega^2+\Omega^2-i2\omega\gamma_{2})-\sigma^2\right].
 \ee
Solving $\tilde{\bf D}^{-1} (\bar{\Omega}_{a,\,\pm})=0$, we find
\bea
&&\bar{\Omega}_{1,\pm}\simeq\Omega_+\pm\frac{i}{2}(\gamma_1+\gamma_2),\\
&&\bar{\Omega}_{2,\pm}\simeq\Omega_-\pm\frac{i}{2}(\gamma_1+\gamma_2),\label{freq}
\eea
with $\Omega_+\equiv\sqrt{\Omega^2+\sigma}$, $\Omega_-\equiv\sqrt{\Omega^2-\sigma}$.

We consider that  two oscillators undergo underdamped oscillations
{in the case of $\Omega^2 > \sigma$ and also both
$\Omega$ and  $\sqrt{\sigma}$ being much larger than $\gamma_1$ and
$\gamma_2$.} Asymptotically as $t \rightarrow \infty$, the
homogeneous solution $\chi_{a;h}$  damps out exponentially and
its contributions to the
covariance matrix with the initial condition dependence can be ignored. Thus, the late-time behavior of
the covariance matrix is determined by a
{particular} solution $\chi_{a;p}$ due to the noise terms $\eta_{T_a}$, which then lead to the
following non-vanishing elements
\begin{align}
 V_{11}&=\frac{1}{2}\langle{\chi_1(\infty),\chi_1(\infty)}\rangle\nonumber\\
 &=\frac{1}{m^2}\int_{-\infty}^{\infty}\frac{d\omega}{2\pi}\left(\lvert\tilde{D}_{11}(\omega)\, \rvert^2 \, \tilde{G}_{H,T_1}(\omega)\, +\lvert\tilde{D}_{12}(\omega)\rvert^2\, \tilde{G}_{H,{T_2}}(\omega)\right),\label{v_11}\\
  V_{22}&=\frac{1}{2}\langle{p_1(\infty),p_1(\infty)}\rangle\nonumber\\
 &=\int_{-\infty}^{\infty}\frac{d\omega}{2\pi}\,\omega^2\, \left(\lvert\tilde{D}_{11}(\omega)\rvert^2\, \tilde{G}_{H, {T_1}}(\omega)\, +\lvert\tilde{D}_{12}(\omega)\rvert^2\, \tilde{G}_{H, {T_2}}(\omega)\right),\label{v_22}\\
  V_{33}&=\frac{1}{2}\langle{\chi_2(\infty),\chi_2(\infty)}\rangle\nonumber\\
 &=\frac{1}{m^2}\int_{-\infty}^{\infty}\frac{d\omega}{2\pi} \, \left(\lvert\tilde{D}_{21}(\omega)\rvert^2\, \tilde{G}_{H,{T_1}}(\omega)\, +\lvert\tilde{D}_{22}(\omega)\rvert^2\, \tilde{G}_{H,{T_2}}(\omega)\right),\label{v_33}\\
  V_{44}&=\frac{1}{2}\langle{p_2(\infty),p_2(\infty)}\rangle\nonumber\\
 &=\int_{-\infty}^{\infty}\frac{d\omega}{2\pi}\,\omega^2 \, \left(\lvert\tilde{D}_{21}(\omega)\rvert^2\, \tilde{G}_{H,{T_1}}(\omega)\, +\lvert\tilde{D}_{22}(\omega)\rvert^2\, \tilde{G}_{H,{T_2}}(\omega)\right),\label{v_44}\\
  V_{13}&=\frac{1}{2}\langle{\chi_1(\infty),\chi_2(\infty)}\rangle\nonumber\\
 &=\frac{1}{m^2}\int_{-\infty}^{\infty}\frac{d\omega}{2\pi}\,
 \left(\tilde{D}_{11}^*(\omega)\tilde{D}_{21}(\omega)\, \tilde{G}_{H,{T_1}}(\omega)\, +\tilde{D}_{12}^*(\omega)\tilde{D}_{22}(\omega)\, \tilde{G}_{H,{T_2}}(\omega)\right),\label{v_13}\\
 V_{14}&=\frac{1}{2}\langle{\chi_1(\infty),p_2(\infty)}\rangle\nonumber\\
 &=\frac{-i}{m}\int_{-\infty}^{\infty}\frac{d\omega}{2\pi}\,\omega\,
 \left(\tilde{D}_{11}^*(\omega)\tilde{D}_{21}(\omega)\, \tilde{G}_{H,{T_1}}(\omega)\, +\tilde{D}_{12}^*(\omega)\tilde{D}_{22}(\omega)\, \tilde{G}_{H,{T_2}}(\omega)\right),\label{v_14}\\
  V_{24}&=\frac{1}{2}\langle{p_1(\infty),p_2(\infty)}\rangle\nonumber\\
 &=\int_{-\infty}^{\infty}\frac{d\omega}{2\pi}\,\omega^2\, \left(\tilde{D}_{11}^*(\omega)\tilde{D}_{21}(\omega)\,
 \tilde{G}_{H,{T_1}}(\omega)\, +\tilde{D}_{12}^*(\omega)\tilde{D}_{22}(\omega)\, \tilde{G}_{H,{T_2}}(\omega)\right)\,\label{v_24} .
\end{align}

In the case of continuous Gaussian variables, it has been shown
that PHS separability criterion in (\ref{PHS}) in quantum systems
 {together with
the} Heisenberg uncertainty relation in (\ref{Hei}) is equivalent
to the following
{inequalities, expressed} in terms of the $Sp(2,R)
\otimes Sp(2,R)$ invariants constructed out of the matrices
$A$,$B$, and $C$ in~(\ref{ABC}) as~\cite{Simon}
\begin{align}
\zeta_\pm=\det{A}\cdot\det{B}-\text{Tr}(A\cdot J\cdot C\cdot J\cdot B\cdot J \cdot C^T\cdot J)-\frac{1}{4}(\det A+\det B) +(\det C \pm\frac{1}{4})^2 \ge 0 \, . \label{zeta_pm}
\end{align}
where the matrix $J$ is defined to be
\begin{align}
J=\begin{pmatrix} 0 & 1\\ -1& 0
\end{pmatrix}\, .
\end{align}
In this form, the condition for $\zeta_-$ is the Heisenberg
uncertainty relation while the condition for $\zeta_+$ is the PHS
criterion, as can be noticed from (\ref{Hei}) and (\ref{PHS}).  In
fact, the change from the density matrix $\rho_{\chi}$ to the
partial transposed density matrix $\bar \rho_{\chi}$ gives the
sign flip in $\det C$. Because $\zeta_+=\zeta_-+ \det C$, the
assumption that the uncertainty relation, $\zeta_-\ge0$ always
true leads to $\zeta_+\ge 0$ when $\det C>0$. For $\det C<0$, the
possibility of $\zeta_+ <0$ implies the  existence of quantum
entanglement~\cite{Simon}. So, the condition, $\det
C=V_{13}V_{24}+V_{14}^2<0$, indicates the possible existence of
quantum entanglement in the {bipartite} Gaussian states, which highlights the
importance of cross-correlations on the entanglements, giving
$\zeta_+ <0$.

Once we have the covariance matrix for the coupled
oscillators in  nonequilibrium steady states, we
 construct $\zeta_+$ according to (\ref{zeta_pm})
\cite{Hsiang2,Hsiang3}. In fact,  ${\bf
\bar V}$ can be diagonalized by the sympletic transformation $S
\in Sp(4,R)$. In particular, the sympletic eigenvalues of partial
transport ${\bf \bar V}$ are
\begin{align}\label{eta_gl}
\bar \eta_{\lessgtr}=\left[\frac{\bar\Delta}{2}\pm \sqrt{\ \frac{\bar \Delta^2}{4}-\det V}\,\,
\right]^{1/2}
\end{align}
where $\bar\Delta =\det A+\det B-2 \det C$.
 Moreover, inequality
(\ref{PHS}) can be recast as a constraint  $\bar \Delta \le
\frac{1}{4}+ 4 \det V $, leading to $ (\bar\eta^2_<
-\frac{1}{4})(\bar\eta^2_>-\frac{1}{4}) \ge 0$. Assuming $\bar
\eta_> >\bar \eta_<$, the PHS separability criterion then simply
reads
  \be
  \label{PHS1}
  \bar \eta_< \ge \frac{1}{2} \,.
  \ee
Violating the separability criterion, namely $\bar \eta_< <
\frac{1}{2}$, can serve as a measure for the entanglement that we
will focus on in this paper. {It has also been suggested that
negativity~\cite{Vid} and logarithmic negativity~\cite{Ple}, which
are defined respectively as,
 \be
 N(\rho)=\max\{0,\frac{1-\bar \eta_<}{2\bar \eta_<}\},~~E(\rho)=\max\{0,-\ln 2\bar \eta_<\}
 \ee
are good entanglement measures. In particular, they do not increase
on average under various local quantum operations. The roles of
these entanglement measures in our setup deserve further study.}

\section{Numerical/analytical analysis of quantum entanglement}\label{sec4}

\begin{figure}[h]
\centering
\includegraphics[scale=1]
{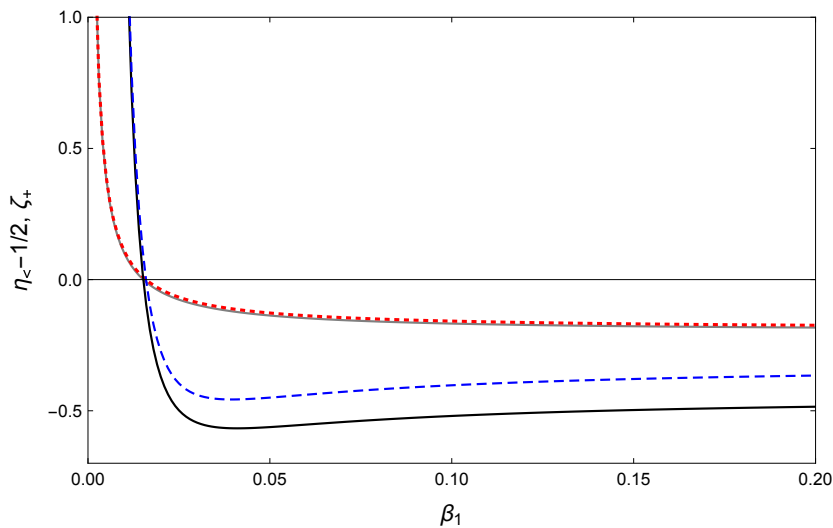} \caption{The separability criterions are all dimensionless and
 depend on  dimensional variables given with respect to some relevant unit of, say $\Omega_0$.
 Thus, the criterions $\zeta_{+}$ (thick
solid line) and $\bar\eta_{<}-1/2$ (thin solid line) are plotted as the function of $\beta_1$ in the unit of $1/\Omega_0$ by fixing $\beta_2=1.5/ \Omega_0$, and  the parameters
$\Omega=5 \Omega_0$, and $\sigma=24 \Omega_0^2$, $\gamma_1=0.005 \Omega_0$, and $\gamma_2=0.25 \Omega_0$
according to the exactly numerical results obtained from (\ref{zeta_pm}) and (\ref{eta_gl})
respectively in that the momentum integration has logarithmic
divergence to be cut off by the chosen cutoff scale $\Lambda=5000\Omega_0$. The approximate
results for $\zeta_{+}$ (blue dotted line) and $\bar\eta_{<}-1/2$ (red
dotted line) are also plotted by substituting  analytically approximate
expressions of  the covariance matrix elements in
(\ref{V_11})-(\ref{V_24}) to (\ref{zeta_pm}) and (\ref{eta_gl}).
We see a  good agreement between the exact and approximate results
in the  parameter regime of validity of the approximations, namely $\beta_1 \Omega_+ \ll1$ and $\beta_2 \Omega_- \gg1$  with $\Omega_+\simeq 7 \Omega_0$ and $\Omega_-\simeq1 \Omega_0$}.
\label{zeta_eta}
\end{figure}

\begin{figure}[h]
\centering
\includegraphics[scale=1.0]
{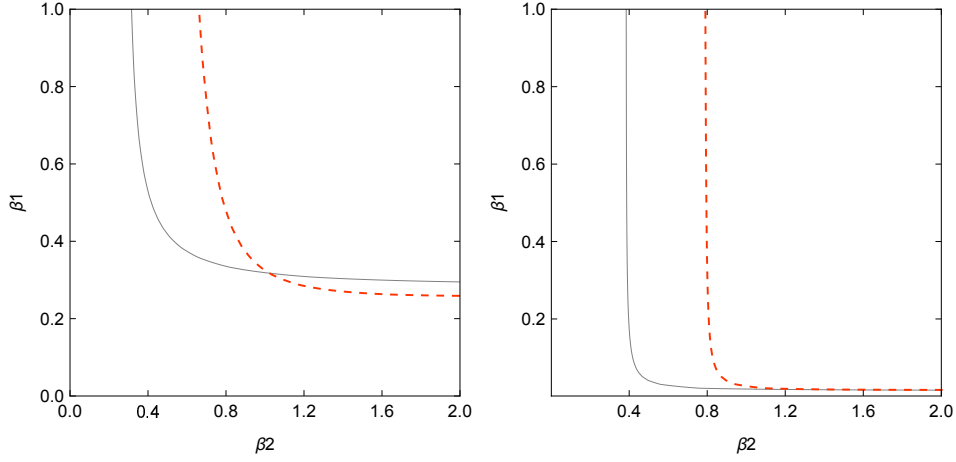} \caption{The line of the critical values of
$\beta_{1c}$ and $\beta_{2c}$ in the unit of $1/\Omega_0$ (black solid line) exactly determined by the
sympletic eigenvalue $\eta_<=1/2$ in (\ref{eta_gl}) and obtained numerically
 is drawn for  $\gamma_1=\gamma_2=0.25\Omega_0$ (left-panel) and  for $\gamma_1=0.005\Omega_0,\,\gamma_2=0.25\Omega_0$ (right-panel) as a comparison where the parameters $\Omega$ and $\sigma$ remain the same as in Fig.~\ref{zeta_eta}. The critical value line  (red dotted line) is also drawn with the same parameters above, but is determined by substituting the approximate  covariance matrix elements in (\ref{V_11})-(\ref{V_24}) into (\ref{eta_gl}), which also show a good agreement with the exact results in the regime of $\beta_1 \Omega_+ \ll1$, $\beta_2 \Omega_- \gg1$  with $\Omega_+\simeq 7 \Omega_0$ and $\Omega_-\simeq1 \Omega_0$}.\label{critical_T}
\end{figure}

Violation of the separability criterions, namely $\zeta_{+} < 0$
and $\bar\eta_< < 1/2$,  indicates the existence of quantum
entanglement of the systems. We now numerically compute $\zeta_+$
and $\bar \eta_{<}$ in~(\ref{zeta_pm}) and (\ref{eta_gl})
respectively to examine their behaviors, which are shown in
Fig.~\ref{zeta_eta}. For a given value $\Omega$, $\sigma$
($\Omega^2 >\sigma$), $\gamma_1$ and $\gamma_2$, and also for a
fixed $\beta_2$, we depict $\zeta_+$ and $\bar \eta_{<}-1/2$ as a
function of $\beta_1$. We find {the} critical value $\beta_{1c}$
obtained from $\zeta_+ = 0$ and $\bar \eta_{<}= 1/2$, above which
($\beta_1 \ge \beta_{1c}$) quantum entanglement exists. In the
range of $\beta_1 \ge \beta_{1c}$, we see that $\bar \eta_{<}$ is
a monotonically decreasing function of $\beta_1$  whereas $\zeta_+
$ is not. It is known that $\bar \eta_{<}$  can serve as a
sensible measure as quantifying the degree of the entanglement
through { negativity and logarithmic negativity for example.}
Fig.~\ref{critical_T} shows the line of the critical values of
$\beta_{1c}$ and $\beta_{2c}$. In the left-panel we choose
$\gamma_1=\gamma_2$, and in the right-panel,$\gamma_1 < \gamma_2$.
In the regime of $\beta_{2c} > \beta_{1c}$ and for the same
$\beta_{2c}$, the corresponding $\beta_{1c}$ given by this
critical temperature line in the case of  $\gamma_1 < \gamma_2$
(right-panel) changes to a much smaller value  than that in the
case of $\gamma_1=\gamma_2$ (left-panel). Nevertheless, for
$\beta_{1c} > \beta_{2c}$, and for the same $\beta_{1c}$, the
value of $\beta_{2c}$  in the case of $\gamma_1 < \gamma_2$
(right-panel) becomes larger  than that in the case
$\gamma_1=\gamma_2$ (left-panel). Thus, the difference between
$\gamma_1$ and $\gamma_2$ ($ \gamma_1 <\gamma_2$) can push the
critical temperature in the bath 1 ($T_{1c}=1/\beta_{1c}$) to a
higher temperature for a fixed $\beta_{2c}$ in the regime
$\beta_{2c}>\beta_{1c}$ whereas it also pulls the critical
temperature ($T_{2c}=1/\beta_{2c}$) in the bath 2 down to a lower
value for a fixed $\beta_{1c}$ in the regime $\beta_{1c} >
\beta_{2c}$. This is one of our main results in this paper to be
further studied analytically in below.

In this two oscillators system, two real parts of (\ref{freq}) are $\Omega_+$ and $\Omega_-$ with $\Omega_+ > \Omega_-$ for $\sigma>0$.
To find the analytical expression for the critical temperatures,
we approximate the covariance matrix elements by considering the high-T limit $\beta_1 \Omega_{+} \ll1$
with reference to the higher frequency mode in the bath 1 and the low-T limit  $\beta_2\Omega_{-} \gg1$
with reference to the lower frequency mode instead in the bath 2, and also
$\gamma_a \ll \Omega_{\pm}$ in the weak damping limit.
Retaining the terms at the relevant order they are summarized as
\begin{align}
&V_{11}=V_{33}\nonumber\\
&\quad\,\,=\frac{1}{m}\left(\frac{\Omega^2}{\Omega_+^2\Omega_-^2}\right)\frac{ \gamma_1}{(\gamma_1+\gamma_2)}\frac{1}{ \beta_1 }+{
\frac{\gamma _2}{4m \left(\gamma _1+\gamma _2\right)} \left(\frac{1}{\Omega_+}+\frac{1}{\Omega_-}\right)}+\frac{\gamma_2}{2\pi m\sigma}\ln \left(\frac{\Omega_+}{\Omega_-}\right)
\nonumber\\
&\quad\quad \,\,-\frac{\gamma _2}{2 \pi  m}\left(\frac{\Omega^2}{\Omega_+^2\Omega_-^2}\right)+\frac{2 \pi  \gamma _2 \sigma ^2}{ 3m{\Omega_+^4\Omega_-^4}}\frac{1}{\beta _2^2}+\mathcal{O}(\gamma_a^2), \label{V_11}
\\[11pt]
&V_{22}=\frac{\gamma_{1}m}{\pi} \ln (\beta_1 \Lambda )\, \Theta (\beta_1 \Lambda -1)+ \frac{m\gamma_1}{(\gamma_1+\gamma_2)}\frac{1}{\beta_1}+{\frac{m\gamma _2}{4 \left(\gamma _1+\gamma _2\right)}\left(\Omega_++\Omega_-\right)}-\frac{m\gamma_2}{2\pi}\nonumber\\
&\quad\qquad+\frac{m\gamma_2\Omega^2}{2\pi\sigma}\ln \left(\frac{\Omega_+}{\Omega_-}\right)+\frac{4 \pi ^3m \gamma _2 \sigma ^2}{15 \Omega_+^4\Omega_-^4}\frac{1}{\beta _2^4}+\mathcal{O}(\gamma_a^2), \label{V_22}
\\[11pt]
&V_{44}=\frac{m\gamma_{1}}{\pi} \ln (\beta_1 \Lambda )\,\Theta (\beta_1 \Lambda -1)+ \frac{m\gamma_1}{(\gamma_1+\gamma_2)}\frac{1}{\beta_1}+\frac{2m\gamma_2}{\pi}\ln\left({\frac{\Lambda}{\Omega_+\Omega_-}}\right)+{\frac{m\gamma _2}{4 \left(\gamma _1+\gamma _2\right)}\left(\Omega_++\Omega_-\right)}\nonumber\\
&\quad\qquad+\frac{m\gamma_2\Omega^2}{2\pi\sigma}\ln \left(\frac{\Omega_+}{\Omega_-}\right)-\frac{m\gamma_2}{2\pi}+\frac{4 \pi ^3m \gamma _2 \sigma ^2}{15 \Omega_+^4\Omega_-^4}\frac{1}{\beta _2^4 }+\mathcal{O}(\gamma_a^2), \label{V_44}
\\[11pt]
&V_{13}=-\frac{1}{m}\left(\frac{\sigma}{\Omega_+^2\Omega_-^2}\right)\frac{\gamma_1}{(\gamma_1+\gamma_2)}\frac{1}{\beta_1}+\frac{\gamma _2  }{4 m\left(\gamma _1+\gamma _2\right)}\left(\frac{1}{\Omega_+}-\frac{1}{\Omega_-}\right)+\frac{\gamma_2 }{2\pi m}\left(\frac{\sigma}{\Omega_+^2\Omega_-^2}\right)\nonumber\\
&\quad\qquad-\frac{2 \pi  \gamma_2 \sigma  \Omega ^2}{3 m\Omega_+^4\Omega_-^4}\frac{1}{\beta _2^2 }+\mathcal{O}(\gamma_a^2),\label{V_13} \\[11pt]
&V_{14}=-\left(\frac{2}{\sigma}\right)\frac{\gamma_1 \gamma_2}{ (\gamma_1+\gamma_2) }\frac{1}{\beta_1}+\frac{8 \pi ^3 \gamma _1 \gamma _2 \sigma }{15 \Omega_+^4\Omega_-^4}\frac{1}{\beta _2^4}+\mathcal{O}(\gamma_a^3),\label{V_14}\\[11pt]
&V_{24}=\left(\frac{m\sigma}{12}\right)\frac{  \gamma_1\beta_1 }{\gamma_1+\gamma_2}+\frac{m\gamma _2 }{4 \left(\gamma _1+\gamma _2\right)}\left(\Omega_+-\Omega_-\right)-\frac{m\gamma_2}{2 \pi}\ln \left(\frac{\Omega_+}{\Omega_-}\right)-\frac{4 \pi ^3 m\gamma _2 \sigma  \Omega ^2}{15 \Omega_+^4\Omega_-^4}\frac{1}{\beta _2^4}+\mathcal{O}(\gamma_a^2). \label{V_24}
\end{align}
\begingroup
Substituting the analytical approximate expressions to
(\ref{zeta_pm}) and (\ref{eta_gl}), we can then compare the approximate results of $\zeta_{+}$
and $\bar\eta_< $ with
their exact numerical ones as shown in Fig.~\ref{zeta_eta},
where the critical value of $\beta_{1c}$ can be correctly obtained
in the case of $\beta_{1c }\Omega_{-} \ll1$ by choosing $\beta_{2c}
\Omega_+ \gg1$ within the  parameter regime where the approximate expressions
are valid. The line of the critical values, $\beta_{1c}$ and
$\beta_{2c}$, resulting from the approximate results are also shown in
Fig.~\ref{critical_T}, which provide a good analytical estimate on
$\beta_{1c}$ and $\beta_{2c}$  again in the regime of $\beta_{1c}
\Omega_{+} \ll1$, $\beta_{2c} \Omega_{-} \gg1$. Substituting
(\ref{V_11})-(\ref{V_24}) into (\ref{zeta_pm}) and (\ref{eta_gl}),
the condition $\bar{\eta}_{<}=1/2$ in the PHS criterion determines
the line of the critical values analytically. To obtain the
analytical expression for the critical temperatures needs further
approximations. Considering the temperature-dependent parts in
(\ref{V_11})-(\ref{V_24}), in the case $\beta_{1c} \Omega_{+} \ll1$,
$\beta_{2c} \Omega_{-} \gg1$, the terms of $1/\beta_2^2$,
$1/\beta_2^4$, and $\beta_1$ are relatively small and thus can be
safely dropped out.  Thus, we consider the  $1/\beta_1$
dependence only.
As for the temperature-independent parts, in the limits of
$\gamma_1 \ll \gamma_2$ and $\gamma_1, \gamma_2 \ll \Omega_{\pm}$,   the
terms proportional to $\gamma$ are ignorably  small except for the
cut-off dependent contributions where the cutoff scale is chosen to be
$\Lambda \gg \Omega$. The PHS criterion then gives
\begin{align}
\beta_{1c} \simeq & 4\pi\gamma_1\bigg/\bigg\{\Big[\pi^2\left(4\gamma_1^2\Omega_+^2+8\gamma_1\gamma_2\Omega_+^2+\gamma_2^2(5\Omega_+^2-2\Omega_+\Omega_-+\Omega_-^2)\right)+\nonumber\\
&\gamma_2^2(\gamma_1+\gamma_2)\ln{\Big\vert\Lambda^4/\Omega_-^4\Big\vert}\left(2\pi(\Omega_--\Omega_+)+(\gamma_1+\gamma_2)\ln{\Big\vert\Lambda^4/\Omega_-^4\Big\vert}\right)\bigg]^{1/2}\nonumber\\
&-\gamma_2(\gamma_1+\gamma_2)\ln{\Big\vert\Lambda^4/\Omega_-^4\Big\vert}-\pi\gamma_2(\Omega_++\Omega_-)\bigg\}, \nonumber\\
\simeq & \frac{4\pi\gamma_1}{\gamma_2\left[\pi\sqrt{5\Omega_+^2-2\Omega_+\Omega_-+\Omega_-^2}-\pi(\Omega_++\Omega_-)-\gamma_2\ln{\Big\vert\Lambda^4/\Omega_-^4\Big\vert}\right]} +\mathcal{O}(\gamma_1^2).
\label{beta1c}
\end{align}
Also, notice that
to be consistent with the small $\gamma$ approximation, the
choice of the cutoff scale is such that the contributions from the
$\ln \Lambda$ dependence are sub-leading as compared with the
leading order terms given by  $\Omega$ and $\sigma$ in~({\ref{beta1c}). The
solution of   $\beta_{1c}$  is found to be $\beta_{1c} \propto
\gamma_1/\gamma_2$ for $\gamma_1 < \gamma_2$ where $\beta_{1c}$ is
significantly suppressed by the smallness of $\gamma_1$ so that
the associated critical temperature $T_{1c}=1/\beta_{1c}$ is boosted into
the hight-$T$ regime with $T_{1c}/\Omega_{+} \gg 1$ with respect to the higher frequency $\Omega_+$ of the normal mode.  On the contrary, in the
regime $\beta_{1c} > \beta_{2c} $, although in Fig.~\ref{critical_T},
$\beta_{2c}$ is not within high-$T$ regime, roughly speaking we expect
$\beta_{2c}$ is given by (\ref{beta1c}) by the replacement of
$\gamma_1 \leftrightarrow \gamma_2$  for $\beta_{1c}$ being a large
value. Thus, for the same $\beta_{1c}$, $\beta_{2c}$ increases  as
$\gamma_1$ decreases from $\gamma_1=\gamma_2$ to $\gamma_1 < \gamma_2$ with a fixed $\gamma_2$ being consistent with the numerical results shown in Fig.~\ref{critical_T}.
In the case of $\gamma_1=\gamma_2$, both $\beta_{1c}$ and
$\beta_{2c}  \propto \mathcal{O}(1/\Omega_+)$ reduces to the findings
in \cite{Hsiang3}. Although the effects
from  the different damping parameters $\gamma_1 \neq \gamma_2$
can push one of the critical temperatures to a higher one, it
may not be so robust to boost the critical temperatures in
both baths to the one higher than the oscillator frequency as in
the case via parametric driving due to the time-dependent
mutual interaction between two oscillators. Our results are consistent with the main conclusion in \cite{Boy} that the large temperature difference in $T_1=1/\beta_1$ and $T_2=1/\beta_2$ in the case of nondegenerate normal modes ($\Omega_+\neq \Omega_-$) can produce the nonvanishing steady-state coherence in high-$T$ limits. And we furthermore show the line of the critical values of $T_{1c}$ and $T_{2c}$ below which the entanglement can survive.
Here we carry out more complete calculations with as few approximation as possible, with which to explore how to achieve hot entanglement between two oscillators.

\section{Comparison to the systems with respective temperature-dependent damping parameters from heat baths}
In this section we  consider the damping
parameter that has the
temperature dependence, say $\gamma_{a T} = \bar\gamma_a T^{\alpha}$ with
$\alpha >0$, resulting from the fact that the system-bath coupling
term is beyond the linear dependence of the bath's
variable.
 In \cite{Lee1,Lee2,Lee3}, we  study the nonequilibrium dynamics of the mirror with
perfect reflection moving in a quantum field.
 The force acting
on the mirror is the radiation pressure of the environmental
field given by  the area integral of the stress tensor, which is quadratic in field variables.
For the system of the $2$-dimensional mirror influenced by a heat bath of free relativistic
scalar fields, the temperature-dependent damping parameters are obtained giving $\alpha=4$.
Later, our studies are extended to  the case of $n$-dimensional mirrors interacting with  a heat bath of Lifshitz scalar fields  with a dynamic exponent
$z$, which are of strongly coupled quantum fields.  The corresponding $\alpha$ using the holographic approach can be computed to be $\alpha=(n+2)/z$.
Taking the strongly self-coupling effects into account, the value of $\alpha$ can even be a fractional number.
With the direct replacement of $\gamma_a \rightarrow \gamma_{aT}=\bar\gamma_a T_a^{\alpha}$ in the Hadamard function $G_{H, T_a}$ (\ref{G_H}) and the matrix $\tilde D$ (\ref{tildeD}), we can numerically calculate the elements of the matrix $V$ from (\ref{v_11})-(\ref{v_24}). Then, from them  the line of the critical values of
$\beta_{1c}$ and $\beta_{2c}$ can be determined by the criterion $\bar{\eta}_{<}=1/2$ in (\ref{eta_gl}), which are shown in Fig.(\ref{fig3}) by choosing the different values of $\alpha$.
In the left-panel (right-panel) of the figure, $\bar \gamma_1 = \bar \gamma_2$ ($\bar \gamma_1 < \bar\gamma_2$) is chosen together with the same values of $\Omega_0$ and $\sigma$ as in Fig.(\ref{zeta_eta}). As $\alpha$ increases while keeping the same $\beta_{2c}$, the corresponding $\beta_{1c}$ increases.
This feature can be  understood  qualitatively from (\ref{beta1c}) for $\beta_{2c} \Omega_- \gg 1$, although the values of $\beta_{1c}$ ($\beta_{1c}\Omega_+ \ge 1$) are beyond the validity of the approximations to achieve it.
With $\gamma_{2T}=\bar\gamma_2/\beta_2^{\alpha}$, as $\alpha$ ($\alpha>0$) increases, for $\beta_{2c} \Omega_- \gg 1$  $\gamma_{2T}$ becomes smaller, leading to the larger values of $\beta_{1c}$.
 In particular, for $\bar \gamma_{2} > \bar\gamma_{1}$, the critical temperature $T_{1c}$ can be boosted into a higher value seen in the right-panel of the figure, which can be qualitatively understandable also from  (\ref{beta1c}) with $\beta_{1c} \propto (\bar \gamma_1/\bar \gamma_2)^{\frac{1}{1+\alpha}}$. {Even for the different $\alpha_a$, namely $\alpha_1\neq \alpha_2$ for the bath 1 and the bath 2, $\beta_{1c} \propto (\bar \gamma_1/\bar \gamma_2)^{\frac{1}{1+\alpha_1}}$.
 \begin{figure}[t]
\centering
\includegraphics[scale=0.6]
{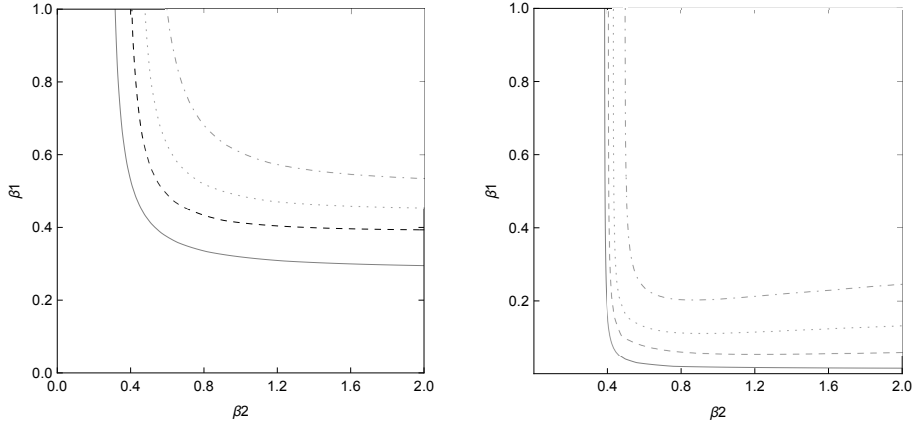} \caption{The line of the critical values of
$\beta_{1c}$ and $\beta_{2c}$ in the unit of $1/\Omega_0$ exactly determined by $\eta_<=1/2$ in (\ref{eta_gl}) with $ \alpha = 0$ (solid), $\alpha = 1/2$
(dashed), $\alpha  = 1$ (dotted) and $\alpha = 2$ (dash-dotted). The values of $\bar\gamma_a$ are chosen to be
$\bar\gamma_{1T}=\bar\gamma_{2T}=0.25\Omega_0^{1-\alpha}$ (left-panel) and  $\bar\gamma_{1T}=0.005\Omega_0^{1-\alpha},\,\bar\gamma_{2T}=0.25\Omega_0^{1-\alpha}$ (right-panel). The parameters $\Omega$ and $\sigma$ remain the same as in Fig.~\ref{zeta_eta}.}\label{fig3}
\end{figure}
 Thus, the suppression of $\beta_{1c}$ is the most significant when $\alpha=0$ also seen in the figure.
Similar behavior of $\beta_{2c}$ for the same $\beta_{1c}$ can also be interpreted from (\ref{beta1c}) by the
exchange of the subscripts $1\leftrightarrow 2$. Thus, in this model, we find that the systems, which couple to heat baths with the temperature independent damping parameters, are more probable to establish hot entanglement by the mutual interaction as compared with the situations with temperature-dependent damping parameters.

\section{Summaries and looking ahead}\label{sec5}

The main goal of this work is to study the entanglement of quantum
systems at finite temperature. We reconsider the model of two
quantum oscillators coupled to its own environment fields through
the coupling linear in the field variable at different temperatures $T_1$ and $T_2$. By
tracing out the environmental degrees of freedom exactly, the
stochastic effective action is obtained, from which the
Heisenberg-Langevin equations for  two
oscillators are derived including the damping effects of the ohmic form  with the
general damping parameters, $\gamma_1 \neq \gamma_2$  as well as
the noise terms where they obey the fluctuation-dissipation relations. Solving the equations for the general solutions allows us
to compute the position and the momentum uncertainties as well as
the expectation values of position-momentum cross correlations of
the system with the given initially separable Gaussian states of
two oscillators. The interaction between two oscillators
start to build up the entanglement between them.  In the weak
damping limit, the late-time behavior of the system is reached by
the nonequilibrium steady state. The separability criterions
constructed by the covariance matrix elements are computed
numerically and analytically. Violation of the  criterions
implies the existence of the entanglement where the
line of the critical temperature can be determined. Considering $\gamma_1
<\gamma_2$, the critical temperature of $T_{1c}$ can be as high as
$T_{1c}/\Omega_+ \propto \gamma_2/\gamma_1$ with reference  to the higher frequency of the normal mode.
 So, we can
have $T_{1c} \gg \Omega_+$ particularly for $\gamma_1 \ll \gamma_2$,
while keeping $T_{2c} \propto \Omega_+$.  This interesting finding
can be compared with the case of $\gamma_1=\gamma_2$ where both
$T_{1c}$ and $T_{2c}$ are found to be the order of $\Omega_+$. Thus, the effects
from  the different damping parameters $\gamma_1 \neq \gamma_2$
can push one of the critical temperatures to a higher one. This might give the possibility of hot entanglement.
Nevertheless, this
may not be so effective to boost the critical temperatures for
both baths as in the case via parametric driving due to the time-dependent interaction between two oscillators. Also we consider the systems with the temperature-dependent damping parameters of the form $\gamma_{a T} = \bar\gamma_a T^{\alpha}$ ($\alpha >0$), which can obtained by coupling to environment fields at finite temperature  via the area integral
of the stress tensor of the field variables. We find that the boost of the critical temperature to a higher value is more noticeable in the case of $\alpha=0$. These findings deserve
further experimental justification. Finally, it might deserve a further study on parametrically driven, dissipative harmonic oscillators with different damping parameters for more exploration on hot entanglement, as an extension of the work in \cite{Gal}.

\section*{Acknowledgments}
 This work was supported in part by the
Ministry of Science and Technology, Taiwan. We are so grateful to  Jen-Tsung Hsiang for illuminating discussions.

\section*{References}
\bibliographystyle{iopart-num}
\bibliography{ref_hot_entanglement}

\end{document}